\documentclass[doublespacing]{elsart}

\usepackage{natbib}

\usepackage{graphics}

\usepackage{amssymb}

\begin{document}

\begin{frontmatter}



\title{Nucleosynthesis of $\bf \rm ^{60}Fe$ in massive stars}


\author[label1,label2]{Marco Limongi}

\address[label1]{INAF - Osservatorio Astronomico di Roma, Via Frascati 33, I-00040, Monteporzio Catone, Rome, Italy}
\address[label2]{Centre for Stellar and Planetary Astrophysics, Monash Unversity, Australia}
\ead{marco@oa-roma.inaf.it}
\ead[url]{http://www.mporzio.astro.it/$\sim$limongi/}

\author[label3,label4]{Alessandro Chieffi}

\address[label3]{INAF - Istituto di Astrofisica Spaziale e Fisica Cosmica, Via Fosso del Cavaliere, Rome, Italy}
\address[label4]{Centre for Stellar and Planetary Astrophysics, Monash Unversity, Australia}
\ead{achieffi@rm.iasf.cnr.it}

\begin{abstract}
We discuss certain aspects of the production  
of $\rm ^{60}Fe$ in massive stars in 
the range between 11 and 120 $\rm M_\odot$, both in the hydrostatic and explosive stages.
We also compare the $\rm ^{60}Fe/^{26}Al$ $\gamma$-ray line flux ratio obtained in the
present calculations to the detected value reported by INTEGRAL/SPI.
\end{abstract}


\end{frontmatter}

\section{Introduction}
\label{intro}
The nucleus $\rm ^{60}Fe$ is a long lived ($\tau\sim$ 2 Myr) radioactive isotope
that should be present in an appreciable amount in our galaxy.
Historically, $\rm ^{60}Fe$ has been considered as a key isotope to understand 
whether or not massive stars are the main contributors to the diffuse
$\rm ^{26}Al$ present in the Galaxy, another long lived radioactive isotope ($\tau\sim$ 1 Myr)
traced by the detected 1.809 MeV $\gamma$-ray emission line at a level
of $\rm \sim 4\times 10^{-4}$ $\rm cm^{-2}$ $\rm s^{-1}$.
Indeed, already in the early 80's \cite{clayton82} pointed out that SNII are the
only candidate sources for $\rm ^{26}Al$ to produce also $\rm ^{60}Fe$ and
hence that the detection of this isotope in the Milky Way could constitute
a strong argument in favour of SNII as the main contributors to the galactic
$\rm ^{26}Al$.
The first detection of $\rm ^{60}Fe$ in the Galaxy was obtained with RHESSI
and reported by \cite{smith03}. The line flux detected implies a 
$\rm ^{60}Fe/^{26}Al$ $\gamma$-ray line flux ratio of $\sim 0.16$ (for each
$\rm ^{60}Fe$ line). More recently \cite{harrisetal05} reported the first detection
of $\rm ^{60}Fe$ decay lines at 1.173 MeV and 1.333 MeV in spectra taken by the SPI
spectrometer on board INTEGRAL during its first year, yielding
a $\gamma$-ray line flux of $\rm 3.7\pm 1.1~\times 10^{-5}~\gamma~cm^{-2}~s^{-1}$.
The same analysis applied to the 1.809 MeV line of $\rm ^{26}Al$ yielded a 
$\rm ^{60}Fe/^{26}Al$ $\gamma$-ray line flux ratio of $\sim 0.11\pm0.03$.
From a theoretical side, many groups have performed calculations of nucleosynthesis
in massive stars, estimating the amounts of either $\rm ^{26}Al$
or both $\rm ^{26}Al$ and $\rm ^{60}Fe$ ejected in the interstellar medium.
However, no set of models covers an extended grid of stellar masses.
Indeed, several research groups interested in
the presupernova evolution and explosion of massive stars provided yields
of both $\rm ^{26}Al$ and $\rm ^{60}Fe$ for stars up to a mass of $\rm 40~M_\odot$
\citep{CL04,RHHW02,WW95,TNH96} - among these works only \cite{RHHW02} included mass loss
in the computations.
On the other hand, groups mainly interested
in the evolution of massive stars including mass loss computed the evolution up to
the end of central He burning and hence provide only the hydrostatic yield
of $\rm ^{26}Al$ \citep{MAPP97,PMVKSCM05,LBF95}. 

In this paper we will discuss to some extent the production of $\rm ^{60}Fe$ in massive stars in 
the range between 11 and 120 $\rm M_\odot$, both in the hydrostatic and explosive stages.
We will also provide theoretical predictions for the 
$\rm ^{60}Fe/^{26}Al$ $\gamma$-ray line flux ratio of such a generation of massive stars and
we will compare them with the observations.

\section{The stellar models}
\label{models}
The yields of $\rm ^{26}Al$ and $\rm ^{60}Fe$ discussed in this paper are based on a new set of presupernova
models and explosions of solar metallicity stars, with mass loss, in the mass range between 11 and 120 $\rm M_\odot$,
covering therefore the full range of masses that are expected to give rise to Type II/Ib/Ic supernovae
as well as those contributing to the Wolf-Rayet populations.
All these models, computed by means of the latest version of the FRANEC code, will be
presented and discussed in a forthcoming paper (Limongi \& Chieffi in preparation).

\begin{figure}[ht]
\begin{center}
\includegraphics{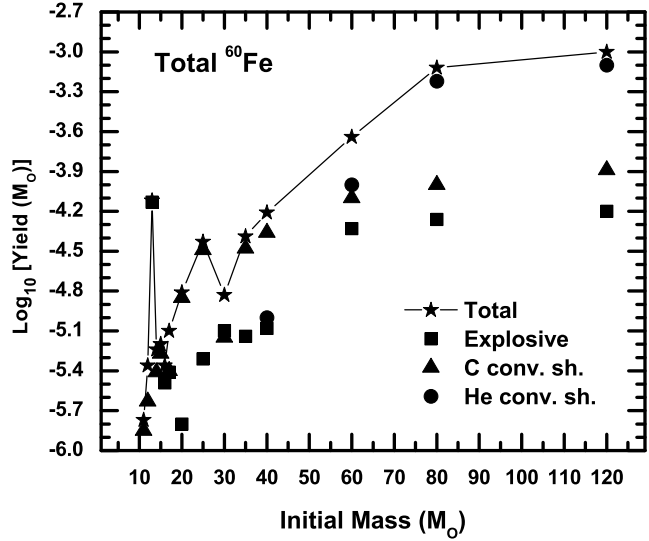}
\caption{Total yield of $\rm ^{60}Fe$ as a function of the initial mass ({\rm filled stars}): the hydrostatic shell
He burning contribution ({\rm filled circles}); the hydrostatic shell C burning contribution ({\rm filled triangles});
the explosive contribution ({\rm filled squares}).}
\end{center}
\end{figure}

\section{The production of $\rm ^{60}Fe$ in massive stars}
\label{fe60}
The isotope $\rm ^{60}Fe$ is an unstable nucleus (its terrestrial decay time is $\rm \tau\simeq 2\times 10^6 y$) that 
lies slightly out of the stability valley, its closest stable neighbor being $\rm ^{58}Fe$. 
At temperatures below $\rm 2\cdot 10^{9}~K$ $\rm ^{60}Fe$ is mainly produced 
by neutron capture on the unstable nucleus $\rm ^{59}Fe$ and destroyed by the $\rm (n,\gamma)$ 
reaction (that always overcomes the beta decay).
Since $\rm ^{59}Fe$ is unstable, the $\rm ^{60}Fe$ production rate depends on the competition
between the $\rm ^{59}Fe(n,\gamma)^{60}Fe$ reaction and the $\rm ^{59}Fe(\beta^{-})^{59}Co$ decay.
At temperatures above $\rm 2\cdot 10^{9}~K$, $\rm ^{60}Fe$
is totally destroyed mainly by $\rm (\gamma,n)$ photodisintegrations and $\rm (p,n)$ reactions.
Hence, two requirements must be fulfilled in order to obtain a substantial production of $\rm ^{60}Fe$:
1) the neutron density must be high enough (i.e. the temperature must be high enough) to allow the
$\rm ^{59}Fe(n,\gamma)^{60}Fe$ reaction to overcome the $\rm ^{59}Fe(\beta^{-})^{59}Co$ decay; 2) the
temperature must be not too high to lead to the complete destruction of $\rm ^{60}Fe$ via $\rm (\gamma,n)$ 
and $\rm (p,n)$ reactions. An order 
of magnitude estimate of the neutron density necessary to cross the  $\rm ^{59}Fe$ bottleneck may be derived 
by equating the (n,$\gamma$) and $\beta^{-}$ decay rates.
A comparison between such a neutron density and the central neutron density obtained during the evolution
shows that no substantial production of $\rm ^{60}Fe$ is obtained in any central burning of a massive
star. Indeed, for temperatures below $\rm 2\cdot 10^{9}~K$ the actual neutron density is always below the
one required to produce a substantial amount of $\rm ^{60}Fe$.
On the contrary, the larger burning temperatures at which the shell burnings occur allow a much higher
production of $\rm ^{60}Fe$. In particular, in stars in the mass interval 40-120 $\rm M_\odot$, shell He
burning occurs at temperatures as high as $\rm 4\cdot 10^{8}~K$. This implies a neutron density
of $\rm 6\times10^{10}-10^{12}~n/cm^3$, the neutrons being produced mainly by the $\rm ^{22}Ne(\alpha,n)^{25}Mg$ 
reaction, and hence a large amount of $\rm ^{60}Fe$. In analogy with shell
He burning, also shell C burning occurs at temperatures high enough ($\rm T \ge 1.3\cdot 10^{9}~K$ for
stars in the mass interval 20-120 $\rm M_\odot$) that a high neutron density is obtained 
($\rm 6\times10^{11}-6\times10^{12}~n/cm^3$) and hence a large amount of $\rm ^{60}Fe$ is synthesized.
Also in this case the main neutron source is the $\rm ^{22}Ne(\alpha,n)^{25}Mg$ reaction, the alpha
particles being provided by the $\rm ^{12}C(^{12}C,\alpha)^{20}Ne$ reaction.
It must be noted, at this point, that the presence of a convective shell plays a crucial role 
for the synthesis of $\rm ^{60}Fe$. Indeed, it has the double responsibility of bringing new fuel 
($\alpha$ particles and $\rm ^{22}Ne$) in the region where the active burning occurs and simultaneously of 
bringing the freshly made $\rm ^{60}Fe$ outward in mass, i.e. at lower temperatures where the neutron density 
becomes negligible and the $\rm ^{60}Fe$ half life increases substantially. Shell Ne burning occurs at temperatures high
enough, but still below the critical value for the total destruction of $\rm ^{60}Fe$, to allow
a large neutron density. However, the lack of an extended 
and stable convective shell lasting up to the moment of the explosion prevents the build up of a significant 
amount of $\rm ^{60}Fe$. In shell O and Si burnings the temperature is so high (above $\rm 2\cdot 10^{9}~K$) that no appreciable
amount of $\rm ^{60}Fe$ is produced.
The hydrostatic production of $\rm ^{60}Fe$ as a function of the stellar mass is shown in Figure 1. A comparison
between the contribution due to the shell He burning and shell C burning shows that in stars below $\rm 60~M_\odot$
the hydrostatic production of $\rm ^{60}Fe$ is dominated by the contribution of the shell C burning while
above this limit the $\rm ^{60}Fe$ is mainly due to the shell He burning. The local minimum corresponding to
the $\rm 30~M_\odot$ is the consequence of the lack of an efficient C convective shell lasting up to the presupernova
stage in this model.
The isotope $\rm ^{60}Fe$ is eventually produced during the explosion in the regions heated up to a temperature
of $\rm T_{peak}\simeq 2.2\times10^{9}~K$. Indeed, for the typical explosive burning timescales ($\sim 1$ s),
below this critical temperature the neutron density is too low to allow a substantial production of 
$\rm ^{60}Fe$ while above this limit $\rm ^{60}Fe$ is totally destroyed by photodisintegration reactions.
Since in most cases such a 
temperature is reached either at the base or within the C convective shell and 
since the matter behind the shock wave can be assumed radiation dominated \citep{ww80},
the amount of  
$\rm ^{60}Fe$ produced during the explosion will depend on the local abundances of $\rm ^{20}Ne$, 
$\rm ^{12}C$, $\rm ^{23}Na$, and  
$\rm ^{22}Ne$ left by the last C convective shell episode and on the final MR relation at the moment of the 
core collapse \citep{CL02}.
Figure 1 shows the total yield of $\rm ^{60}Fe$ as a function of the stellar mass
together with the explosive and hydrostatic contributions. There is a global direct scaling with 
the initial mass and a quite monotonic behavior: the two exceptions are the $\rm 13~M_\odot$ and the $\rm 30~M_\odot$ models.
The $\rm 13~M_\odot$ model constitutes a striking exception because a large amount of $\rm ^{60}Fe$
is synthesized by the explosion in this case. The reason is that the peak temperature of $\rm 
2.2\times10^9$ K occurs beyond the outer border of the C convective shell where the abundances of 
$\rm ^{12}C$ and $\rm ^{22}Ne$, in particular, are much higher than in the C convective shell.
The minimum corresponding to the $\rm 30~M_\odot$ model has been already discussed (see above). 
Below $\rm 60~M_\odot$ the total yield is dominated by the contribution of the C 
convective shell while above this mass it is the He convective shell to play the major role. The explosive 
burning almost always plays a minor role. 

\begin{figure}[ht]
\begin{center}
\includegraphics{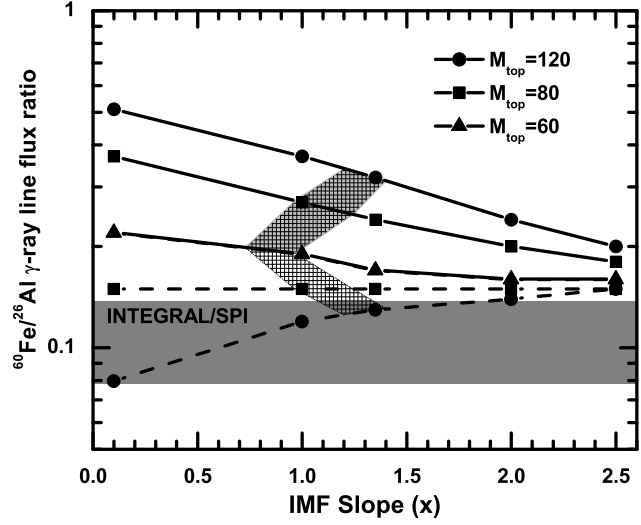}
\caption{$\rm ^{60}Fe/^{26}Al$ $\gamma$-ray line flux ratio integrated over a single power low IMF
as a function of the slope of the IMF for three different upper mass limits; 
$\rm M_{top}=60~M_\odot$ ({\rm filled triangles});
$\rm M_{top}=80~M_\odot$ ({\rm filled squares});
$\rm M_{top}=120~M_\odot$ ({\rm filled circles}). The solid lines refer to the standard models,
while the dashed lines to models in which the contribution to the $\rm ^{60}Fe$ of the 
He convective shell of the more massive stars is removed.}
\end{center}
\end{figure}

\section{Comparison to observations}
\label{compobs}
To compare properly with the observations (see introduction), the yields of both $\rm ^{26}Al$ and $\rm ^{60}Fe$ 
must be integrated over a stellar Initial Mass Function (IMF) $\phi(m)$.
We assume here that the IMF is described by a single power low, i.e., $\phi(m)=km^{-1+x}$.
Figure 2 shows the $\rm ^{60}Fe/^{26}Al$ $\gamma$-ray line flux ratio as a function of
the slope $x$ of the IMF, obtained for three different values of the IMF upper mass limit,
i.e., $\rm M_{top}=60,~80$ and $\rm 120~M_\odot$ (solid lines). 
The horizontal shaded area refers to the $\rm ^{60}Fe/^{26}Al$ $\gamma$-ray line flux ratio reported
by INTEGRAL/SPI \citep{harrisetal05} while the hatched areas correspond to the
region where the ratio between Type Ib/c and Type II supernovae is compatible with the
observed value of $0.3\pm 0.04$, as reported by \citep{ct01}. Figure 2 clearly shows that
our theoretical predictions always overestimate the observed 
$\rm ^{60}Fe/^{26}Al$ $\gamma$-ray line flux ratio for any choice of the slope of the IMF
and the IMF upper mass limit. This is due to the copious $\rm ^{60}Fe$ production in stars more
massive than $\rm 40~M_\odot$. Indeed, 
the fit is much more improved if such a contribution 
is artificially removed in these stars (dashed lines in Figure 2). 
A more detailed discussion of the implications of these
result will be addressed soon in a forthcoming paper.





\begin{thebibliography}{}


\harvarditem{Cappellaro \& Turatto}{2001}{ct01} Cappellaro, E., and Turatto, M. 2001, ASSL Vol.~264, "The Influence of Binaries on Stellar Population Studies," 199 
\harvarditem{Chieffi \& Limongi}{2002}{CL02} Chieffi, A., and Limongi, M. 2002, ApJ, 577, 281
\harvarditem{Chieffi \& Limongi}{2004}{CL04} Chieffi, A., and Limongi, M. 2004, ApJ, 608, 405
\harvarditem{Clayton}{1982}{clayton82} Clayton, D.D. 1982, in "Essays in Nuclear Astrophysics", ed. C.Barnes et al. (Cambridge:CUP), 401 
\harvarditem{Harris et al.}{2005}{harrisetal05} Harris, M.J., Kn\"odlseder, J., Jean, P., Cisana, E., Diehl, R., Lichti, G.G., Roques, J.P., Schanne, S., and Weidenspointner, G. 2005, A\&A, 433, L49
\harvarditem{Langer et al.}{1995}{LBF95} Langer, N., Braun, H., and Fliegner, J. 1995, Ap\&SS, 224, 275
\harvarditem{Meynet et al.}{1997}{MAPP97} Meynet, G., Arnould, M., Prantzos, N., and Paulus, G. 1997, A\&A, 320, 460
\harvarditem{Palacios et al.}{2005}{PMVKSCM05} Palacios, A., Meynet, G., Vuissoz, C., Kn\"odlseder, J., Schaerer, D., Cervi\~no, M., and Molawi, N. 2005, A\&A, 429, 613
\harvarditem{Rauscher et al.}{2002}{RHHW02} Rauscher, T., Heger, A., Hoffman, R., and Woosley, S.E. 2002, ApJ, 574, 323
\harvarditem{Smith}{2003}{smith03} Smith, D.M. 2003, ApJL, 589, L55 
\harvarditem{Thielemann et al.}{1996}{TNH96} Thielemann, F.K., Nomoto, K., and Hashimoto, M. 1996, ApJ, 460, 408
\harvarditem{Weaver et al.}{1978}{WZW78} Weaver, T.A., Zimmerman, G.B., and Woosley, S.E. 1978, ApJ, 225, 1021
\harvarditem{Weaver \& Woosley}{1980}{ww80} Weaver, T.A., and Woosley, S.E. 1980, in Ann. NY Acad. Sci., 336, Ninth Texas Symp. on Relativistic Astrophysics, ed. J. Ehlers, J. Pressy, \& M. Walker, 335
\harvarditem{Woosley \& Weaver}{1995}{WW95} Woosley, S.E., and Weaver, T.A. 1995, ApJS, 101, 181

\end{thebibliography}
\end{document}